\documentstyle[aps,eqsecnum,multicol,graphicx]{revtex}

\newcommand{\bra}[1]{\langle {#1} |}
\newcommand{\ket}[1]{| {#1} \rangle}

\newcommand{\vecr}{{\bf r}}

\begin{document}

\draft
\title{Photoabsorption spectra in the continuum
of molecules and atomic clusters}
\author{Takashi~Nakatsukasa\footnote{
       Email: Ntakashi@riken.go.jp}}
\address{RI Beam Science Laboratory, RIKEN, 2-1 Hirosawa,
Wako 351-0198, Japan}
\author{
Kazuhiro~Yabana\footnote{
       Email: yabana@nucl.ph.tsukuba.ac.jp}}
\address{Institute of Physics, University of Tsukuba,
Tennodai 1-1-1, Tsukuba 305-8571, Japan.}

\maketitle

\tighten
\begin{abstract}
We present linear response theories in the continuum capable of
describing photoionization spectra and dynamic
polarizabilities of finite systems with no spatial symmetry.
Our formulations are based on the time-dependent local density 
approximation with uniform grid representation in the three-dimensional 
Cartesian coordinate.
Effects of the continuum are taken into account either with a Green's
function method or with a complex absorbing potential in a real-time method.
The two methods are applied to a negatively charged cluster in the 
spherical jellium model and to
some small molecules (silane, acetylene and ethylene).
\end{abstract}

\pacs{PACS number(s): 31.15.Ew, 31.70.Hq, 33.80.Eh}


\begin{multicols}{2}
\section{Introduction}

Oscillator strength distribution characterizes the optical response of 
atoms and molecules. Advances in measurements with synchrotron radiation 
and high resolution electron energy loss spectroscopy have enabled us to 
obtain oscillator strength distribution of a whole spectral region 
originated from valence electrons \cite{Hat99}. The photon energy 
dependences of molecular photoelectron spectra have also been measured.
They provide useful information about the dynamics of the photoionization 
processes.

Theoretical analysis of photoabsorption spectra above the first ionization 
threshold requires continuum electronic wave function in a non-spherical 
multi-center potential. Several methods have been developed for this 
purpose\cite{CCRM91}, including the continuum multiple scattering 
method\cite{DD74}, the Schwinger variational method\cite{LRM82,BL88}, 
finite-volume variational method\cite{LR84}, the linear algebraic 
method\cite{CS84}, and the $K$-matrix method\cite{CCM85}. The Stieltjes 
moment method does not directly utilize the scattering state but 
extract continuum spectrum from the spectral moments which are
calculated with a square-integrable basis set\cite{Lan80}. 
The method has been extensively applied to large systems\cite{CYA97,PPA97}. 
The complex basis method also gives continuum spectra by the calculation
with a square-integrable basis set\cite{YPM85}.

The purpose of the present article is to present alternative theoretical 
methods for the continuum spectra of molecules. Our methods rely
upon the linear response theory in the time-dependent local-density 
approximation (TDLDA) and employ a uniform grid representation in a 
three-dimensional Cartesian coordinate.

The TDLDA (or alternatively called the time-dependent density-functional
theory) is an extension of the static density-functional theory 
to describe electronic dynamics under a time-dependent external 
field\cite{RG84}. In the practical applications, an adiabatic approximation 
is usually assumed: the same exchange-correlation potential as 
that in the static case is used for the time-dependent problem. 
The correlation effect is included through the dynamical screening which
is represented by an induced local potential.
In the TDLDA, the continuum boundary condition was treated with the radial 
Green's function for spherical systems such as atoms\cite{ZS80} and 
clusters in the spherical jellium model\cite{Eka84,Ber90}. 
The importance of the dynamical screening effect on the continuum response 
has been stressed. The method has been extended for molecules with
a single-center expansion technique\cite{LS84}. However, the application
was limited to small, axially symmetric molecules.

Since the Hamiltonian in the Kohn-Sham theory is almost diagonal in 
coordinate representation, a grid representation in the coordinate space 
provides an economical description. A uniform grid representation in the 
three-dimensional Cartesian coordinate which has been developed by 
Chelikowsky et.al\cite{Che94} provides a convenient basis for our purpose.
The problem is then how to incorporate the scattering boundary condition 
in the uniform grid representation.

Our first method is based on the real-time method which one of the present 
authors has recently developed to calculate dynamic polarizability
of finite systems\cite{YB96,YB99}. In the method, the time-dependent 
Kohn-Sham equation is solved explicitly in real-time as well as in 
real-space with a uniform grid. The dynamic polarizability of a whole 
spectral region is then obtained at once through the time-frequency 
Fourier transformation. 
An advantage of this method is that we do not 
need to handle matrices of very large dimensionality, since only 
a single Slater determinant is evolved in time.
A drawback is that the single-particle continuum states cannot be 
treated exactly because the electrons are confined in a limited size 
of box (model space). A possible way to avoid the difficulty would 
be employing a complex absorbing potential at the end of the box,
which we shall discuss later.

Our second method is based on the modified Sternheimer method\cite{Mah80}.
The method has been widely used in the linear response calculations 
\cite{LA91,IYB00}. It recasts the response problem into solving the 
static Schr\"odinger-type equation with a source term. The problem is 
then to solve the equation with an appropriate outgoing boundary condition.
Our recipe here is to solve iteratively the equation taking into account
the boundary condition employing a Green's function of a spherical potential, 
separating the self-consistent potential into long range spherical and 
short range non-spherical parts.

The paper is organized as follows. We present our methods in
Sec. \ref{sec: formalism}.
The real-time method with an absorptive boundary potential and the Green's
function method are explained. In Sec. \ref{sec: applications},
applications of the methods will be demonstrated.
First the spherical jellium cluster is considered to 
confirm reliability of our methods. We then show calculations of 
photoabsorption spectra of some small molecules and compare them with 
measurements. We give interpretation of the obtained spectra.
Finally conclusion are drawn in Sec. \ref{sec: conclusions}.

\section{Linear response in the continuum of three-dimensional real space}
\label{sec: formalism}

\subsection{Real-time method with an absorbing potential}

In this section we first recapitulate the real-time method in 
TDLDA \cite{YB96,Cal00}. For systems with relatively large number 
of particles, the real-time method is one of the most efficient 
method to calculate the electronic excitations in molecules and 
atomic clusters. 

The TDLDA equations for a spin-independent $N$-electron system are given
in terms of the time-dependent Hamiltonian
$h(t)$ which is a functional of the density $n(\vecr,t)$.
\begin{eqnarray}
\label{TDLDA}
&&i\frac{\partial}{\partial t} \phi_i(\vecr,t) = h(t) \phi_i(\vecr,t),
  \quad i=1,\cdots,N/2 \\
\label{h_sp}
&&h(t) = -\frac{1}{2m}\nabla^2
+ V_{\rm ion} \nonumber \\
&& \hspace{2cm}+ e^2 \int d^3 r' \frac{n(\vecr',t)}{|\vecr - \vecr'|}
+\mu_{\rm xc}[n(\vecr,t)] ,\\
&&n(\vecr,t) = 2 \sum_{i=1}^{N/2} \left| \phi_i(\vecr,t) \right|^2 .
\end{eqnarray}
We adopt $\hbar=1$ throughout the present paper.
$V_{\rm ion}$ is an electron-ion potential for which we employ
a norm-conserving pseudopotential\cite{TM91} with a separable
approximation\cite{KB82}. $\mu_{\rm xc}$ is an exchange-correlation 
potential. 

We represent the electronic wave function on a uniform grid points
inside a certain spatial area\cite{Che94}. All potentials in 
Eq.~(\ref{h_sp}) are diagonal in this representation except for the 
nonlocal part of $V_{\rm ion}$. The second order differential 
operator in the kinetic energy is approximated employing a 
nine-point formula.
For the time evolution, we use an algorithm developed in Ref.\cite{FKW78}
which utilize a predictor-corrector method.
The integration of Eq.~(\ref{TDLDA}) is approximately carried out with 
a fourth-order Taylor expansion.:
\begin{eqnarray}
\phi_i(t+\Delta t) &=& e^{-ih(t+\Delta t/2)\Delta t} \phi_i(t) \nonumber \\
&\approx& \sum_{n=0}^4 \frac{(-ih(t+\Delta t/2)\Delta t)^n}{n!} \phi_i(t) .
\end{eqnarray}
For stable iteration, the time step $\Delta t$ should be smaller
than the inverse of the maximum eigenvalue of the Hamiltonian. 
The energy and the particle-number conservations are then well satisfied.
This method was successful in nuclear physics to investigate
the dynamics of nuclear reaction \cite{FKW78,Neg82}
and has been proved to be fruitful for electron systems
as well \cite{YB96,Cal00}.

We start with the static LDA problem solving the ground-state Kohn-Sham
equations and determine the initial occupied orbitals $\phi_i(\vecr)$
and density $n_0(\vecr)$. The conjugate gradient method is used to solve
the static Kohn-Sham equation.
Then, an external field is turned on instantaneously at $t=0$.
In this paper, we shall consider only the dipole response, 
adopting $V_{\rm ext}=k_0 r_\nu \delta(t)$ ($\nu=x,y,z$)
which produces an initial state for the time evolution as
\begin{equation}
\label{initial_state}
\phi_i(\vecr,0_+)=e^{-ik_0r_\nu} \phi_i(\vecr) , \quad \nu=x,y,z.
\end{equation}
We calculate time evolution with this initial condition without any 
external fields.
The dynamical polarizabilities in the time representation is then
obtained as
\begin{equation}
\alpha_\nu(t) = -\frac{e^2}{k_0} \int d^3r r_\nu \delta n(\vecr,t) ,
\quad \nu=x,y,z,
\label{alpha_t}
\end{equation}
where the transition density is simply given by the difference
from that of the ground-state
\begin{equation}
\label{delta_n_t}
\delta n(\vecr,t) = n(\vecr,t) - n_0(\vecr) .
\end{equation}
Since the dynamical polarizability is usually
defined in the energy representation,
we take the Fourier transform of Eq.~(\ref{alpha_t}).
\begin{equation}
\label{alpha_omega}
\alpha_\nu(\omega) = \int_0^T dt \alpha_\nu(t) e^{i\omega t-\Gamma t/2} ,
\quad \nu=x,y,z,
\end{equation}
where we introduce a smoothing parameter $\Gamma$.
In order to obtain $\alpha(\omega)$ in energy resolution of $\Gamma$,
we need to calculate the time evolution up to $T \geq 2\pi/\Gamma$.
In the next section, we discuss a gas-phase average of
the $x$, $y$ and $z$ direction, as
\begin{equation}
\alpha(\omega)=1/3 \sum_{\nu=x,y,z} \alpha_\nu(\omega).
\end{equation}

When the energy is in a region where $\omega+\epsilon_i > 0$,
a part of the electron wave functions $\phi_i(\vecr,t)$
can escape to the continuum.
This outgoing part of wave function is eventually reflected at the edge of
the box and produces a spurious discrete structure in the photoabsorption 
spectra. Therefore, we now introduce an imaginary absorbing potential 
$W(\vecr)$ to suppress the reflection and mimic the continuum.
Since the imaginary potential should be zero in a region where
the ground-state density has a finite value,
we should adopt a spherical box of large radius
and the imaginary potential is switched on only in a outer shell
of width $\Delta r$.
Although the complex potential at the edge of the box slightly violates
normalization and energy conservation,
the box must be large enough to preserve the conservations
if the time evolution is carried out using an initial state of Eq.
(\ref{initial_state}) with $k_0=0$ (the ground state).

We adopt an absorptive potential of a linear dependence on coordinate, 
which has been discussed in the wave-packet method for molecular 
collisions\cite{NB89,Chi91}:
\begin{equation}
\label{absorbing_pot}
W(r)=\left\{
\begin{array}{ll}
0 & \mbox{ for } 0< r < R ,\\
-iW_0\frac{r-R}{\Delta r}  & \mbox{ for }  R < r < R + \Delta r.
\end{array}
\right.
\end{equation}
The height $W_0$ and width $\Delta r$ must be carefully chosen
to prevent the reflection. There has been an argument based on the WKB 
theory to elucidate a condition of no reflection \cite{NB89,Chi91}.

For simplicity, let us take a one-dimensional system.
The absorbing potential is given by a form similar to
Eq. (\ref{absorbing_pot}):
$W(x)=0$ for $x < 0$ and $W(x)=-iW_0 x /\Delta x$
for $x > 0$.
The WKB solution is
\begin{eqnarray}
\psi(x)&=&\exp(ikx) + R\exp(-ikx) \quad\mbox{ for } x < 0, \\
\psi(x)&=& (1+R)E^{1/4} \left( E+iW_0\frac{x}{\Delta x}\right)^{-1/4}
 \nonumber \\
 && \hspace*{0.5cm}\times\exp\left\{i\int^x_0 dx k(x)\right\}
 \quad\mbox{ for } x > 0,
\end{eqnarray}
where $k(x)=[2m(E+iW_0x/\Delta x)]^{1/2}$ and $k=k(0)$.
Conditions we require are
$|R|^2 \ll 1$ (no reflection) and $|\psi(\Delta x)| \ll 1$
(complete absorption).
A continuity condition for a derivative at $x=0$ leads to
\begin{equation}
R\approx \frac{W_0}{4\sqrt{8m}E^{3/2}\Delta x} ,
\end{equation}
assuming $W_0/E \ll 1$.
The absolute value of wave function at $x=\Delta x$ is approximately given by
\begin{equation}
\left| \psi(\Delta x)\right| \approx
 \exp\left\{ -W_0 \Delta x \left( \frac{m}{8E}\right)^{1/2}\right\} .
\end{equation}
If we demand $|R|^2 < 0.001$ and $|\psi(\Delta x)|^2 < 0.01$,
we have
\begin{equation}
\label{absorb_cond_1}
20 \frac{E^{1/2}}{\Delta r \sqrt{8m}} < | W_0 |
 < \frac{1}{10} \Delta r \sqrt{8m} E^{3/2} ,
\end{equation}
where $\Delta x$ has been replaced by $\Delta r$.
This is the condition discussed in Ref. \cite{Chi91}
which we shall test with numerical calculations.
Strictly speaking, the conditions,
$|R|^2 < 0.001$ and $|\psi(\Delta x)|^2 < 0.01$,
lead to a factor
$18.4$ instead of $20$ in the left hand side
and $0.128$ instead of $1/10$ in the right hand side.
In actual real-time calculations, electrons with different energies
are emitted simultaneously.
Thus, $\Delta r$ and $W_0$ must be chosen properly according to the
energy spectrum of photoabsorption above the ionization threshold.


\subsection{Green's function method with an outgoing boundary condition}

Exact treatment of the continuum is possible with the use of a 
Green's function. For spherical systems, the Green's function can easily
be constructed by making a multipole expansion and discretizing the
radial coordinate. The method was first applied to nuclear giant 
resonances \cite{BT75,SB75} and then applied to photoabsorption in 
atoms \cite{SZ80,ZS80}. In this section we present a method to construct 
a Green's function in the uniform grid representation for a system 
without any spatial symmetry.

The linear response theory is formulated most conveniently using a
density-density correlation function \cite{FW71}.
The general formalism with local-density approximation is
given in the frequency representation \cite{SB75,ZS80}.
The transition density, which corresponds to the Fourier transform
of Eq. (\ref{delta_n_t}), can be expressed with the use of
the independent-particle density-density correlation function $\chi_0$:
\begin{equation}
\label{delta_n_omega}
\delta n(\vecr,\omega) = \int d^3 r' \chi_0(\vecr,\vecr';\omega) 
V_{\rm scf}(\vecr',\omega) ,
\end{equation}
where $V_{\rm scf}$ is the self-consistent field which is the sum of
the external field and the dynamical screening (dielectric) field:
\begin{equation}
\label{Vscf}
V_{\rm scf}(\vecr,\omega) = V_{\rm ext}(\vecr,\omega)
 + \int d^3r' \left.
         \frac{\delta V[n(\vecr)]}{\delta n(\vecr')}
         \right|_{n=n_0}
         \delta n(\vecr';\omega) ,
\end{equation}
where $V[n(\vecr)]$ is a single-particle potential of
the time-independent version of Eq.~(\ref{h_sp}),
defined by $h=-\nabla^2/2m + V$.
Although a part of $V$ is non-local,
the screening field arises from density-dependent parts of $V$,
the direct Coulomb and the exchange-correlation potentials,
which are local.
The calculation neglecting the second term of Eq.~(\ref{Vscf})
will be called an "independent-particle approximation" (IPA)
in the next section.

The $\chi_0$ has a form
\begin{eqnarray}
\label{chi_0_1}
\chi_0(\vecr,\vecr';\omega) = && 2
 \sum_i^{\rm occ}\sum_m^{\rm unocc} \left\{
 \phi_i(\vecr)\frac{\phi_m^*(\vecr)\phi_m(\vecr')}
              {\epsilon_i-\epsilon_m-\omega-i\eta} \phi_i^*(\vecr')
 \right.\nonumber \\
&& +\left.\phi_i^*(\vecr)\frac{\phi_m(\vecr)\phi_m^*(\vecr')}
              {\epsilon_i-\epsilon_m+\omega+i\eta} \phi_i(\vecr')\right\} ,
\end{eqnarray}
where $\eta$ is an infinitesimal positive parameter.
The factor two in the right hand side of Eq. (\ref{chi_0_1}) is the
spin degeneracy of each single-particle state.
Here the summation with respect to unoccupied states $m$ contains
both discrete and continuum states.
Instead of taking the explicit summation, we may use the single-particle
Green's function defined by
\begin{eqnarray}
\label{G_sp}
G^{(\pm)}(\vecr,\vecr';E)
&=& \bra{\vecr}\left( E-h[n_0] \pm i\eta \right)^{-1}\ket{\vecr'} \nonumber \\
&=& \sum_k^{\rm all}\frac{\phi_k(\vecr)\phi_k^*(\vecr')}{E-\epsilon_k\pm i\eta}.
\end{eqnarray}
The sign determines a boundary condition of the Green's function;
the $(+)$ for a outgoing wave and the $(-)$ for an incoming wave.
In fact, the summation with respect to $m$ in Eq. (\ref{chi_0_1}) can be
extended to all states because the summation over the occupied states
for the first term will be canceled by a contribution from the second term.
Therefore, using Eq.~(\ref{G_sp}), we can rewrite (\ref{chi_0_1}) as
\begin{eqnarray}
\label{chi_0_2}
\chi_0(\vecr,\vecr';\omega) = && 2
 \sum_i^{\rm occ}
 \phi_i(\vecr)\left\{
     \left(G^{(+)}(\vecr,\vecr';(\epsilon_i-\omega)^*)\right)^*
   \right. \nonumber \\
   && +\left. G^{(+)}(\vecr,\vecr';\epsilon_i+\omega) \right\}
   \phi_i(\vecr') ,
\end{eqnarray}
where we assume that the occupied states have real wave functions.

To calculate the dipole photoresponse of a system,
one should take $V_{\rm ext}=r_\nu$, $\nu=x,y,z$.
Then, the dynamic polarizability is given by
\begin{equation}
\alpha_\nu(\omega)=-e^2 \int d^3r r_\nu \delta n(\vecr,\omega) ,
\quad \nu=x,y,z.
\end{equation}

Since the number of spatial grid points is large, it is not convenient 
and even impossible to construct explicitly the response function 
$\chi_0(\vecr,\vecr';\omega)$ and to perform spatial integration by 
summing up over grid points in solving the self-consistent equations, 
Eqs.~(\ref{delta_n_omega}) and (\ref{Vscf}). We can avoid the difficulty
by converting the integral into the equivalent differential equation.
We introduce a function
\begin{equation}
\label{psi_i}
\psi_i(\vecr;E,V_{\rm scf}) \equiv
\int d^3r' G^{(+)}(\vecr,\vecr';E) V_{\rm scf}(\vecr') \phi_i(\vecr') .
\end{equation}
The transition density is then expressed as
\begin{eqnarray}
\label{delta_n_omega_2}
\delta n(\vecr,\omega) = 2 \sum_i^{\rm occ} \phi_i(\vecr) &&
\left\{ \left(\psi_i(\vecr;(\epsilon_i-\omega)^*,V_{\rm scf}^*)\right)^*
 \right. \nonumber \\
&& \quad + \left. \psi_i(\vecr;\epsilon_i+\omega,V_{\rm scf}) \right\} .
\end{eqnarray}
When the energy $\omega$ is
far below the first ionization threshold, all the
outgoing channels are closed and Eq.~(\ref{psi_i}) is equivalent to the
Schr\"odinger equation of energy $E=\epsilon_i\pm\omega$
with a source term,
\begin{equation}
\label{Sternheimer}
(E - h[n_0]) \ket{\psi_i} = V_{\rm scf}\ket{\phi_i} ,
\end{equation}
assuming that the function $\bra{\vecr}\psi_i\rangle $
outside a box area vanishes.
The integral in Eq.~(\ref{psi_i}) is thus converted into a linear
algebraic equation (\ref{Sternheimer}). This procedure is known as the
modified Sternheimer method\cite{Mah80}.
However, since this cannot describe a correct asymptotic behavior
of the wave function,
it is not applicable when the energy $E$ is close to zero.
Furthermore, when $E>0$, the method is incapable to produce the continuum
spectra.

Thus, the remaining task is to calculate $\psi_i(\vecr;E,V_{\rm scf})$ defined
by Eq.~(\ref{psi_i}) under an appropriate outgoing boundary condition.
For this purpose, we employ an integral equation for the Green's function.
We start with a single-particle problem with a spherical potential
$V_0(\vecr)$.
For instance, the Green's function for free particles ($V_0(\vecr)=0$)
has an analytic expression,
\begin{equation}
\label{G_0}
G_0^{(+)}(\vecr,\vecr';E)=-\frac{m}{2\pi} \frac{e^{+ik|\vecr-\vecr'|}}
      {|\vecr-\vecr'|} ,
\end{equation}
where $k=\sqrt{2mE}$ ($E>0$).
For a negative energy $E<0$, $\exp(+ik |\vecr -\vecr'|)$
in Eq.~(\ref{G_0}) is replaced by $\exp(-\kappa |\vecr -\vecr'|)$
with $\kappa=\sqrt{-2mE}$.
The single-particle Green's function for a non-spherical (even non-local)
potential $V(\vecr,\vecr')$ then satisfies the following integral equation
\begin{eqnarray}
\label{Dyson_eq}
&& G^{(+)}(\vecr,\vecr';E) = G_0^{(+)}(\vecr,\vecr';E)
 + \int d^3r^{''} d^3r^{'''} \nonumber \\
&& \hspace{1.5cm} G_0^{(+)}(\vecr,\vecr^{''};E)
               \tilde{V}(\vecr^{''},\vecr^{'''})
               G^{(+)}(\vecr^{'''},\vecr';E) ,
\end{eqnarray}
where
$\tilde{V}(\vecr,\vecr')=V(\vecr,\vecr')-V_0(\vecr)\delta^3(\vecr-\vecr')$.
The boundary condition of $G_0^{(+)}$ determines an asymptotic
behavior of $G^{(+)}$.
Substituting this into Eq.~(\ref{psi_i}), we obtain a multi-linear
equations for $\psi_i(\vecr;E,V_{\rm scf})$,
\begin{eqnarray}
\label{Dyson_eq_2}
 && \psi_i(\vecr;E,V_{\rm scf}) \nonumber \\
 &&\hspace{1cm}
  -\int d^3r' d^3r^{''} G_0^{(+)}(\vecr,\vecr';E)\tilde{V}(\vecr',\vecr^{''})
 \psi_i(\vecr^{''};E,V_{\rm scf}) \nonumber\\
 &&\hspace{1cm}
  =\int d^3r' G_0^{(+)}(\vecr,\vecr';E) V_{\rm scf}(\vecr') \phi_i(\vecr').
\end{eqnarray}
In solving this equation, we need to evaluate the following integral
\begin{equation}
\label{psi_G0}
\Psi(\vecr;E)=\int d^3r' G_0^{(+)}(\vecr,\vecr';E) f(\vecr'),
\end{equation}
where $f(\vecr)$ is either
$\int d^3r' \tilde V(\vecr,\vecr') \psi_i(\vecr';E,V_{\rm scf})$
or $V_{\rm scf}(\vecr) \phi_i(\vecr)$. We note that both of them are zero
outside the box. We calculate $\Psi(\vecr;E)$ again by recasting
Eq.~(\ref{psi_G0}) into equivalent differential equation,
\begin{equation}
\label{dif_eq_G0}
\left\{ E - \left(-\frac{1}{2m}\nabla^2 +V_0(\vecr)\right) \right\} 
\Psi(\vecr;E)  = f(\vecr).
\end{equation}
In the discretization, this is a linear algebraic equation for grid 
points inside the box area. However, the solution outside the box area 
is needed to apply the Laplacian operator and should be prepared by other
method. We prepare it by a multipole expansion method. Noting that 
$V_0(\vecr)$ is a central potential, $G_0^{(+)}$ can be expressed
in terms of the regular and irregular solutions of radial Schr\"odinger
equation in the usual way.
\begin{eqnarray}
\label{Psi}
&& \left.\Psi(\vecr;E)\right|_{r\geq R} =
 \sum_{lm}^{l_{\rm max}} \frac{w_l^{(+)}(r;E)}{r}
  Y_{lm}(\hat r) \Phi_{lm}(E) ,\\
\label{Phi_lm}
&& \Phi_{lm}(E) \equiv 2m \int_{r<R} d^3r'
 \frac{u_l(r';E)}{r'} Y_{lm}(\hat r') f(\vecr'),
\end{eqnarray}
where $u_l(r;E)$ and $w_l^{(+)}(r;E)$ are solutions of the radial
differential equation being normalized as the Wronskian
$W[u_l,w_l^{(+)}]$ is unity.
The $u_l$ is regular at the origin and $w_l^{(+)}$ is an outgoing
wave at infinity.
The $l_{\rm max}=16$ is chosen in the later calculations.

We must choose the $V_0(\vecr)$ so as to make
the potential $\tilde{V}(\vecr)=V(\vecr)-V_0(\vecr)$
be negligible outside the box.
For neutral molecules,
the attractive ionic potential is approximately canceled out
by the repulsive direct Coulomb term.
However we will later employ a gradient-corrected exchange potential 
which possesses a correct asymptotic form of $-e^2/r$.
Therefore we use a jellium potential, Eq.~(\ref{jellium_pot}), with $Z=1$ as 
$V_0(\vecr)$. The $w_l^{(+)}(r;E)$ in Eq. (\ref{Psi}) is an irregular 
Coulomb wave function with an outgoing boundary condition.
We use a FORTRAN77 program ``COULCC'' in Ref. \cite{TB85} to calculate
Coulomb functions of complex energies.
The $u_l(r,E)$ in Eq. (\ref{Phi_lm}) is calculated by integrating the 
radial Schr\"odinger equation with a fourth-order Runge-Kutta method.

We now summarize our procedure to solve the response equation.
Once an external field $V_{\rm ext}(\vecr,\omega)$ is given, 
Equations (\ref{delta_n_omega}) and (\ref{Vscf}) constitute a linear 
equation for the transition density $\delta n(\vecr,\omega)$. 
Discretizing on a uniform grid, we have a linear algebraic equation 
with dimensionality equal to the number of grid points.
In order to solve the equation, we need to calculate
actions of the $\chi_0$ on some functions, which is equivalent to
calculate actions of the $G^{(+)}$.
$\psi_i(\vecr;E,V_{\rm scf})$ is calculated by solving another linear
equation (\ref{Dyson_eq_2}).
In order to solve Eq. (\ref{Dyson_eq_2}), we need to calculate
actions of the $G_0^{(+)}$.
This operation is achieved by solving Eq.~(\ref{dif_eq_G0}), again
discretizing on grid points. We note that the outgoing boundary condition
is imposed at this stage. The wave function $\Psi(\vecr;E)$
outside the box is prepared by a multipole expansion method,
Eqs.~(\ref{Psi}) and (\ref{Phi_lm}).
Solutions of linear equations (\ref{delta_n_omega}), (\ref{Dyson_eq_2})
and (\ref{dif_eq_G0}) are obtained by iterative methods.
The iterative procedure is summarized in Fig. \ref{algorithm}.

Our algorithm thus requires to solve multiple nested 
linear algebraic equations. Conjugate gradient (CG) method and its variants 
offer stable and efficient scheme to solve these equations.
If the energy $E$ is real, Eq. (\ref{dif_eq_G0}) is a hermitian problem.
Therefore, the CG method is very powerful to solve the 
equation. However, if the energy is complex, Eq. (\ref{dif_eq_G0}) becomes
a non-hermitian problem to which the CG method is not applicable.
For such cases, we adopt a Bi-conjugate gradient (Bi-CG) method.
To solve Eq.~(\ref{Dyson_eq_2}), we have tested different iterative 
methods to this non-hermitian problem. We found that a generalized 
conjugate residual (GCR) method is one of the most effective algorithms 
for the present problem.  For the outer most iteration loop of 
Eq. (\ref{delta_n_omega_2}), we use the GCR method again.

\begin{minipage}{0.45\textwidth}
\begin{figure}[ht]
\centerline{\includegraphics[width=0.8\textwidth]{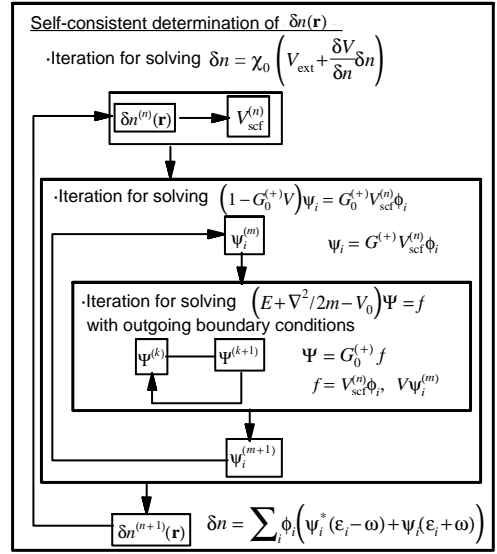}}
\vspace*{0.5cm}
\caption{
Algorithm to determine the transition density
$\delta n(\vecr,\omega)$ is presented.
There are three nested iterative loops to solve linear equations.
}
\label{algorithm}
\end{figure}
\end{minipage}

\section{Applications}
\label{sec: applications}

\subsection{Spherical jellium model for Na$_7^-$: Test study}
\label{sec: Na7-}

First we have carried out a calculation for
a negatively-charged Na cluster Na$_7^-$.
We use a spherical uniformly-charged jellium potential for $V_{\rm ion}(r)$.
The jellium potential of radius $R_I$ is given by
\begin{equation}
\label{jellium_pot}
V_{\rm ion}(r)=\left\{
\begin{array}{ll}
-\frac{Ze^2}{2R_I}\left\{3-\left(\frac{r}{R_I}\right)^2\right\}
 & \mbox { for } r\leq R_I ,\\
-\frac{Ze^2}{r} & \mbox { for } r > R_I ,
\end{array}
\right.
\end{equation}
where $Ze$ is a total charge of a jellium sphere.
For the Na$_7^-$, we have $Z=7$ and eight valence electrons.
The main photoabsorption peak is
calculated to be above the ionization threshold.
Thus, this would be a good test case to study a response in the
continuum and to check the applicability of the theories
in the previous section.

We discretize three-dimensional Cartesian coordinates with
spacing $\Delta x=\Delta y=\Delta z=1.5$ \AA\ and employ grid
points inside a spherical box of radius $R=12$ \AA.
This is found enough for the ground-state density to be
negligible at the edge $r=R$.
The number of mesh points is 2109.
The TDLDA Hamiltonian is given by Eq.~(\ref{h_sp}) in which the
exchange-correlation potential is that of Ref.~\cite{GL76}:
\begin{equation}
\label{mu_xc_GL76}
\mu_{\rm xc}[n(\vecr)] = -\frac{1.222}{r_s(\vecr)}
-0.0666 \ln \left( 1+\frac{11.4}{r_s(\vecr)} \right) ,
\end{equation}
in units of Ry where $4/3 \pi r_s^3=n(\vecr)$.
The radius of jellium sphere $R_{\rm I}$ is $3.93\times 8^{1/3}$ a.u.

Since the jellium potential for Na$_7^-$ is spherical,
we can use the same technique as that of
Ref.~\cite{SB75,ZS80} (a Fortran program in Ref. \cite{Ber90})
and confirm that our methods provide the same results.
The Green's function is expanded in partial waves and is given by
\begin{equation}
\label{G_l}
G_l^{(+)}(r,r';E) = u_l(r_<;E) w_l^{(+)}(r_>;E)/W[u_l,w_l^{(+)}] ,
\end{equation}
for the $l$-th partial wave.
Here $u_l$ and $w_l^{(+)}$ are regular and irregular solutions
of the radial differential equation, respectively.
The boundary condition of $w_l^{(+)}$ at the edge $r=R$ is defined
as $\exp(ikr)$ with $k^2=2m(E-e^2/R)$ for $E>e^2/R$
and $\exp(-\kappa r)$ with $\kappa^2=2m(e^2/R-E)$ for $E<e^2/R$.
Of course, this cause a change of the threshold energy by $e^2/R$,
but it is enough to check validity of our methodology.
We also add a small imaginary parameter $\Gamma/2$ to the frequency $\omega$,
which plays exactly the same role as a smoothing parameter of
the Fourier transform in real-time method, Eq.~(\ref{alpha_omega}).
In this calculation, $\Gamma = 0.1$ eV is used and
the mesh size for radial coordinate is as small as 0.1 \AA.

\begin{minipage}{0.45\textwidth}
\begin{figure}[htb]
\centerline{\includegraphics[width=0.7\textwidth]{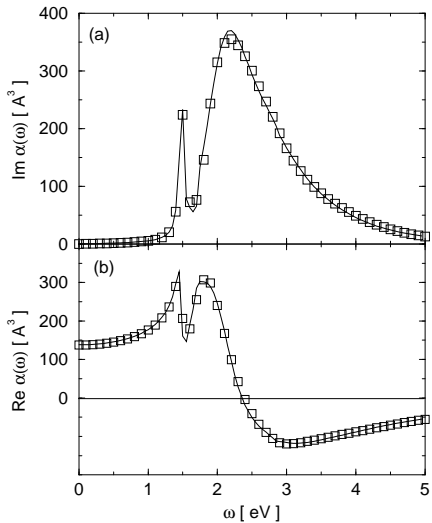}}
\caption{
Imaginary (a) and real (b) part of dynamic polarizabilities
for Na$_7^-$ cluster
calculated with the Green's function method.
Solid lines are the results of one-dimensional calculation of
Ref.~\protect\cite{Ber90} of jellium sphere
and square symbols are those of the three-dimensional
calculation without assuming any symmetry.
The smoothing parameter $\Gamma=0.1$ eV is used for both calculations.
}
\label{Na7-_GF}
\end{figure}
\end{minipage}

For the three-dimensional Green's function calculation,
in order to justify Eq.~(\ref{Dyson_eq}),
a condition that $\tilde{V}(\vecr)\approx 0$ for $r >R$ must be satisfied.
We assume that the asymptotic behavior is
the same as that of the (radial) one-dimensional calculation.
Namely, we adopt the Green's function of free particles, Eq.~(\ref{G_0}),
with an energy being shifted as $E \rightarrow E-e^2/R$.
We also use the same smoothing parameter $\Gamma=0.1$ eV.
It turns out that the inclusion of $\Gamma$ helps convergence of
iterative procedure.

We show the results in Fig.~\ref{Na7-_GF}.
The three-dimensional calculation has been done
for a frequency range $0<\omega<5$ eV with a step $\Delta\omega=0.1$ eV.
The solid lines are the results of one-dimensional calculation in which
the Green's functions are expanded in the partial wave, Eq.~(\ref{G_l}).
The squares are the results of the three-dimensional calculation which
perfectly agree with the solid lines.
This means that the Green's function method on the three-dimensional mesh
is able to take account properly of the continuum effects.

The static calculation indicates the highest occupied molecular orbital
(HOMO) $-0.37$ eV, and a Coulomb potential has a value $e^2/R=1.2$ eV at
the edge of the box $R=12$ \AA.
Thus, the ionization threshold is 1.57 eV in this calculation.
The figure indicates that the photoabsorption has peaks below and
above the threshold.
In the IPA calculation, we obtain a single peak at an energy
1.4 eV.
The screening potential brings the peak into the continuum region
(around 2.35 eV) and splits the single peak into two parts at the threshold.
The summed oscillator strengths for energies up to 5 eV correspond to
about 95\% of the Thomas-Kuhn-Reiche (TRK) sum rule for valence electrons.

\begin{minipage}{0.45\textwidth}
\begin{figure}[htb]
\centerline{\includegraphics[width=0.65\textwidth]{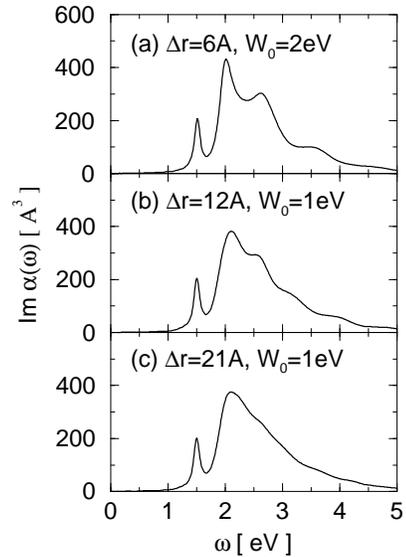}}
\caption{
Imaginary part of dynamic polarizabilities for Na$_7^-$ cluster
calculated with the real-time method.
The absorptive potentials are
chosen as (a) $\Delta r=6$ \AA\ with $W_0=2$ eV, (b) $\Delta r=12$ \AA\ with
$W_0=1$ eV, (c) $\Delta r=21$ \AA\ with $W_0=1$ eV.
The smoothing parameter $\Gamma=0.1$ eV is used on the Fourier transform
of Eq.~(\ref{alpha_omega}).
}
\label{Na7-_RT}
\end{figure}
\end{minipage}

We now turn to the real-time method.
A necessary size of space is much larger than that of the Green's function
method because the absorbing potential is set up in a outer region
$R< r < R+\Delta r$.
We prepare three kinds of absorbing potentials in a linear form of
Eq.~(\ref{absorbing_pot}):
(a) $\Delta r=6$ \AA\ with $W_0=2$ eV,
(b) $\Delta r=12$ \AA\ with $W_0=1$ eV and (c)
$\Delta r=21$ \AA\ with $W_0=1$ eV.
Using a square mesh of $\Delta x=\Delta y=\Delta z=1.5$ \AA,
the numbers of mesh points are (a) 7153, (b) 17077, and (c) 44473,
respectively.
We use a duration of time evolution $T=70$ eV$^{-1}$ with a time step
$\Delta t=0.01$ eV$^{-1}$ and a smoothing parameter $\Gamma=0.1$ eV for
the Fourier transform.
Since we have a complex potential at the edge of the box,
the energy is not strictly conserved
but the scale of violation turns out to be
less than 0.1\% for this calculation.
The imaginary part of dynamic polarizability is shown in Fig.~\ref{Na7-_RT}.
Parts (a), (b) and (c) correspond to the three cases described
above.
For (a) and (b), the main resonance in the continuum region looks like
a superposition of three different peaks
which are located at $\omega\approx 2$, 2.7 and 3.5 eV, respectively.
However, since we cannot see the same behavior in Fig.~\ref{Na7-_GF} (a),
this should be a spurious effect of the reflection of outgoing waves.
This is confirmed by a calculation of (c), for which
the structure at $\omega > 2.5$ eV is almost disappeared and we see
only a smooth tail.

Now let us check the criterion of a good absorber,
Eq. (\ref{absorb_cond_1}).
The position of the main peak is calculated
at $E_{\rm res}-E_{\rm thr}= 2.35-1.57 \approx 0.8$ eV
above the ionization threshold.
Using E=0.8 eV and $(8m)^{-1}\approx 1$ eV \AA$^2$ for electrons,
Eq. (\ref{absorb_cond_1}) reduces to
\begin{equation}
18(\Delta r)^{-1} < | W_0 | < 0.07 \Delta r ,
\end{equation}
where $W_0$ in units of eV and $\Delta r$ in \AA.
This criterion is satisfied for the case (c)
but not for (a) and (b).
Fig.~\ref{Na7-_RT} shows that the real-time method provides us with
a correct response in the continuum if no reflection occurs
at the edge of the box.
In order to find a suitable absorbing potential (of a linear form),
we find the criterion, Eq.~(\ref{absorb_cond_1}), very useful
even when a condition $W_0/E \ll 1$ is not satisfied.

\subsection{Valence shell photoabsorption of silane}
\label{sec: SiH4}

It is well known that the energy of the highest occupied orbital does
not coincide with the first ionization potential in the simplest
local-density approximation. This fact causes a serious  problem
for the continuum response calculation that the ionization threshold 
cannot be adequately described by the static Kohn-Sham Hamiltonian. 
Furthermore, the excited states around the ionization threshold appear 
in too low excitation energies. To remedy this defect, a gradient
correction for the exchange-correlation potential has been proposed.
We utilize the one constructed by van Leeuven and Baerends\cite{LB94} 
which we denote as $\mu^{\rm (LB)}$. It is so constructed that the potential 
has a correct $-e^2/r$ tail asymptotically. The energy of the highest 
occupied orbital also approximately coincides with the ionization 
potential. For small molecules, TDLDA calculations with this gradient 
correction have shown to give accurate description of discrete excitations 
in small molecules\cite{Cas98}.

In the following sections \ref{sec: SiH4}, \ref{sec: C2H2} and 
\ref{sec: C2H4}, we employ a sum of the exchange-correlation potentials 
of $\mu^{\rm (PZ)}$ of Ref. \cite{PZ81} for the local-density part and
$\mu^{\rm (LB)}$ for
the gradient correction.
We should remark here that an accurate calculation of the gradient correction
$\mu^{(\rm LB)}(\vecr)$ becomes difficult at far outside the molecule,
because the $\mu^{(\rm LB)}$ depends on $|\nabla n(\vecr)|/n(\vecr)^{4/3}$
which approaches to finite but both the numerator and the
denominator approaches to zero at $r\rightarrow \infty$.
Thus, we use an explicit asymptotic form,
$\mu^{(\rm LB)}(\vecr)=-e^2/r$ for $r > R_{\rm c}$.
In the followings,
the $R_{\rm c}$ is chosen as 6.5 \AA\ for silane and
5.5 \AA\ for acetylene and ethylene.

When we evaluate the screening field in Eq. (\ref{Vscf}),
we neglect the functional derivative of $\mu^{\rm (LB)}(\vecr)$
with respect to the density because this should be a small correction.
For the real-time calculation, we make the same approximation.
Namely the time-dependent exchange-correlation potential is calculated
by
\begin{eqnarray}
\mu_{\rm xc}(\vecr,t)&\approx&
 \mu_{\rm xc}[n_0(\vecr)] + \int d^3r'
 \left.\frac{\delta \mu_{\rm xc}(\vecr)}{\delta n(\vecr')}\right|_{n=n_0}
  \delta n(\vecr',t) \nonumber \\
 &\approx& \mu_{\rm xc}[n_0(\vecr)] + \int d^3r'
 \left.\frac{\delta \mu^{(\rm PZ)}(\vecr)}{\delta n(\vecr')}\right|_{n=n_0}
  \delta n(\vecr',t) . \nonumber
\end{eqnarray}

In this section, we discuss the application of our techniques
to a deformed molecule, silane SiH$_4$.
We use the pseudopotentials which are calculated by the prescriptions
of Refs.~\cite{TM91,KB82}.
For Si atom, we employ pseudopotentials for $s,p,d$ waves and take
$d$-wave potential as a local part.
For the hydrogen atom, we take $s$ and $p$ potentials with the latter
as a local potential.
The box is a sphere of radius $R=7$ \AA\ discretized in meshes
of $\Delta x=\Delta y=\Delta z=0.4$ \AA.
Position of the Si atom (nucleus) is located at the origin and
four hydrogen atoms are at
$(1.209, 0.0, 0.855)$, $(-1.209, 0.0, 0.855)$,
$(0.0, 1.209, -0.855)$, and $(0.0, -1.209, -0.855)$
in units of \AA.
The symmetry of this molecule belongs to the tetrahedral ($T_d$) point
group. The valence electronic orbitals in the ground state are
$$
\label{T_d}
(3a_1)^2, (2t_2)^6: \ ^1A_1 .
$$
In actual calculations, we do not assume a priori the symmetry, but
the degeneracies of electron orbitals according to Eq. (\ref{T_d})
naturally emerge after the minimization of the total energy.
For this molecule, the dipole response does not depend on the direction 
of an external field. A smoothing parameter $\Gamma=0.5$ eV is used in 
the followings.

Utilizing the exchange-correlation potential of
$\mu^{\rm (PZ)}+\mu^{(\rm LB)}$,
the calculated occupied valence orbitals in the ground state
are listed in Table \ref{IP}.
If we neglect the gradient correction term $\mu^{\rm (LB)}$,
we obtain $-13.5$ eV for $3a_1$ and $-8.3$ eV for $2t_2$, respectively.
The photoabsorption oscillator strengths
for silane calculated by means of the
Green's function method are shown in Fig.~\ref{SiH4_GF}.
The oscillator strength $d f/d\omega$, the photoabsorption cross section
$\sigma(\omega)$, and the imaginary part of dynamic polarizability
${\rm Im}\alpha(\omega)$, are related to each other by
\begin{equation}
\sigma(\omega)=\frac{2\pi^2 e^2}{mc}\frac{d f}{d\omega}
= \frac{4\pi\omega}{c} {\rm Im}\alpha(\omega) .
\end{equation}

A sharp peak at 10 eV in the calculation consists of bound discrete peaks 
overlapped by the width $\Gamma$. In the experiment\cite{Kam91,Coo95,Hat99},
we observe a peak at 10.7 eV, the width of which is considered to originate 
from a coupling of electronic excitation with ionic motion.
Since we fix the ion coordinates and calculate vertical electronic
excitations, we cannot describe the width below the ionization threshold.

On the other hand, a broad peak at 14.6 eV in the experiment
is above the ionization threshold and the main decay channel is 
an emission of the electron\cite{Kam91,Hat99}.
The calculation well reproduces this peak although the energy
shifts slightly to the lower (14 eV).
Beyond the peak of 14.6 eV, a group of tiny peaks is observed in the
experiment, which is assigned the Rydberg states with a hole $(3a_1)^{-1}$
embedded in the continuum.
This fine structure is smeared out in the calculation because
the smoothing parameter $\Gamma$ is much larger than the width of each
Rydberg state.
The integrated oscillator strengths for energies up to 30 eV is calculated
to exhaust 87\% of the TRK sum rule for valence electrons.
About 60\% of the valence shell strengths lie between 10 and 20 eV
(Fig. \ref{SiH4_RT} (b)).
\begin{minipage}{0.45\textwidth}
\begin{figure}[htb]
\centerline{\includegraphics[width=0.9\textwidth]{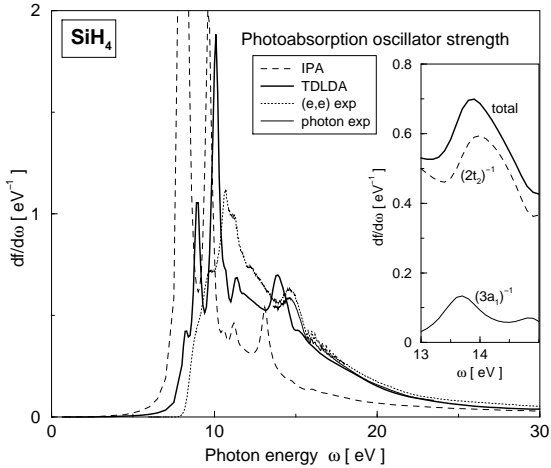}}
\caption{
Calculated and experimental photoabsorption oscillator strengths
of silane.
The thick solid line is the calculation compared with
synchrotron radiation experiment (thin solid line)
\protect\cite{Kam91} and
high resolution dipole (e,e) experiment (dotted line) \protect\cite{Coo95}.
The dashed line is the IPA calculation without dynamical screening.
The smoothing parameter $\Gamma=0.5$ eV is used in the calculation.
Inset: An energy region of $13 < \omega < 15$ eV is magnified and
the total oscillator strengths are decomposed into those associated with
different occupied valence electrons, $3a_1$ (solid line)
and $2t_2$ (dashed line).
}
\label{SiH4_GF}
\end{figure}
\end{minipage}

The screening field in Eq. (\ref{Vscf}) significantly reduces
a dipole polarizability at low frequency.
The calculated static dipole polarizability is 5.1 \AA$^3$
while it is 7.9 \AA$^3$ in the IPA.
The dielectric effects also change the absorption spectra.
The screening shifts the resonance energies to higher energies
because the electron-electron correlation is repulsive.
Roughly speaking, the IPA overestimates the absorption spectra
at low frequency and underestimates at high frequency.
As you see in Fig. \ref{SiH4_GF},
the dielectric effects are very important
to reproduce the magnitude and shape of the photoresponse.

In order to discuss details of each resonance, it is useful to
calculate a partial oscillator strength of each occupied orbital
which should be identified with a photoemission spectra if the
neutral dissociation channel without electron emission is negligibly
small. As we see in Ref. \cite{ZS80},
the photoabsorption cross section can be written as a sum of contributions
from all occupied orbitals. Here we define the partial oscillator strengths
$(df/d\omega)_i$ as
\begin{eqnarray}
&&\frac{df}{d\omega}=
  2 \sum_i^{\rm occ} \left(\frac{df}{d\omega}\right)_i ,\\
&&\left(\frac{df}{d\omega}\right)_i
 = -\frac{2\omega}{\pi e^2} \int d^3r d^3r'
     V_{\rm scf}^*(\vecr) \phi_i(\vecr) \times \nonumber \\
&& \hspace{0.5cm} {\rm Im} 
 \left\{\left(\tilde{G}^{(+)}(\vecr,\vecr';(\epsilon_i-\omega)^*)\right)^*
 +\tilde{G}^{(+)}(\vecr,\vecr';\epsilon_i+\omega)\right\} \nonumber \\
&& \hspace{4cm} \times V_{\rm scf}(\vecr') \phi_i(\vecr'),
\end{eqnarray}
where $\tilde{G}^{(+)}$ is defined by Eq. (\ref{G_sp}) except that the
summation with respect to $k$ is carried out only for unoccupied states.

\begin{minipage}{0.45\textwidth}
\begin{figure}[htb]
\centerline{\includegraphics[width=0.9\textwidth]{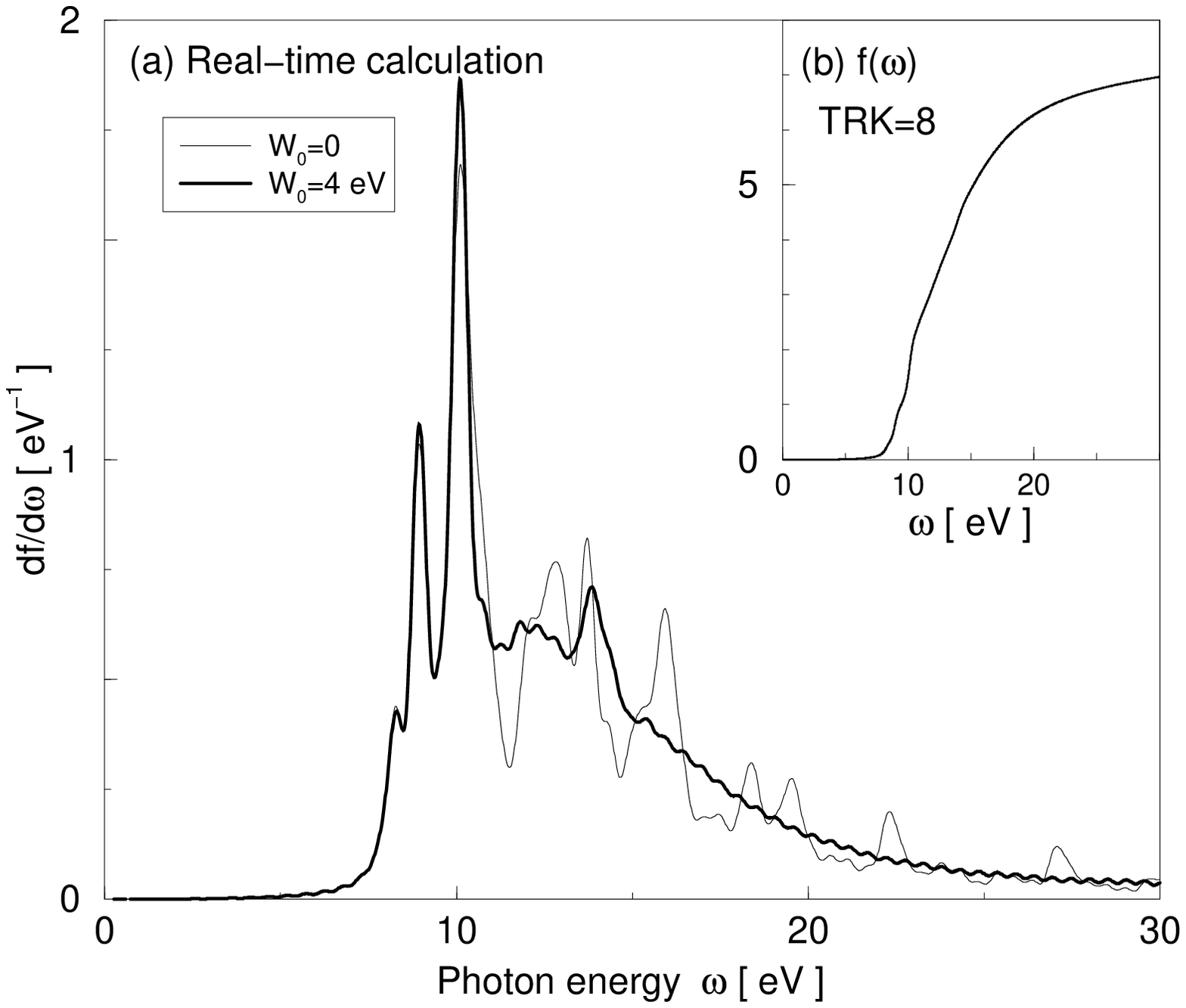}}
\caption{
(a) Photoabsorption oscillator strengths of silane calculated with
the real-time method.
The thin line corresponds to the result obtained with the box of $R=10$ \AA\ 
without a complex absorbing potential, while
the thick line to the one with a complex potential
of $W_0=4$ eV with $\Delta r=10$ \AA\ in addition to a spherical box of
$R=7$ \AA.
In carrying out the Fourier transform, the smoothing parameter $\Gamma=0.5$
eV is used.
(b) The integrated oscillator strength $f(\omega)$.
The TRK sum rule indicates $f(\infty)=8$ for the valence shell photoabsorption.
}
\label{SiH4_RT}
\end{figure}
\end{minipage}
\newpage

\end{multicols}
\centerline{
\begin{minipage}{0.8\textwidth}
\newcommand\s{\ \ \ \ }
\begin{table}[htb]
\caption{
Calculated eigenvalues of occupied valence orbitals and
experimental vertical ionization potential (IP) in units of eV.
The experimental data are taken from Ref. \protect\cite{PP72} for silane,
from Ref. \protect\cite{Uka91} for acetylene and from
Ref. \protect\cite{Hol97} for ethylene.
}
\begin{tabular}{ccccccccc}
\multicolumn{3}{c}{Silane} & \multicolumn{3}{c}{Acetylene}
                        & \multicolumn{3}{c}{Ethylene} \\
occ & cal & exp & occ & cal & exp & occ & cal & exp \\
state &    & IP  & state &    & IP  & state &    & IP  \\
\hline
$(3a_1)^2$ & $-17.4$ & $ 18.2$ & $(2\sigma_g)^2$ & $-22.4$ & $ 23.5$ &
 $(2a_g)^2$ & $-22.8$ & $ 23.7$ \\
$(2t_2)^6$ & $-12.4$ & $ 12.8$ & $(2\sigma_u)^2$ & $-18.4$ & $ 18.4$ &
 $(2b_{1u})^2$ & $-18.6$ & $ 19.4$ \\
         &         &         & $(3\sigma_g)^2$ & $-16.7$ & $ 16.4$ &
 $(1b_{2u})^2$ & $-16.3$ & $ 16.3$ \\
         &         &         & $(1\pi_u)^4$    & $-12.1$ & $ 11.4$ &
 $(3a_g)^2$ & $-14.7$ & $ 14.9$ \\
         &         &         &             &         &         &
 $(1b_{3g})^2$ & $-13.2$ & $ 13.0$ \\
         &         &         &             &         &         &
 $(1b_{3u})^2$ & $-11.7$ & $ 11.0$ \\
\end{tabular}
\label{IP}
\end{table}
\end{minipage}
}
\begin{multicols}{2}

Below the ionization threshold, there are three bound transitions in our
calculations which we find are associated with the excitations of $2t_2$
valence electrons. The first two transitions are considered to correspond
to the shoulders in the measurements\cite{Coo95,Wu93} and interpreted as 
transitions to the Rydberg states.
The broad resonance at 14 eV (14.6 eV in experiment) is above
the ionization threshold of the $2t_2$ orbitals.
Therefore, the structure may be more complex.
An inset of Fig. \ref{SiH4_GF} shows the partial oscillator strengths of
$3a_1$ and $2t_2$ orbitals in a photon-energy region of 13 to 15 eV.
The calculation suggests that the resonance structure is due to
the bound-to-bound excitation of $3a_1$ electrons. The excitation
couples with the bound-to-continuum excitations of $2t_2$ electrons
through the dynamical screening effect. The coupling shifts the excitation
energy up by about 1 eV and brings about the width due to the autoionization
process. The partial cross section of $2t_2$ electrons also acquires
oscillation
as a function of the excitation energy due to the rapid change of the induced 
field.

Now let us examine the applicability of the real-time method.
The time step is chosen as $\Delta t=0.002$ eV$^{-1}$ and the time evolution
is calculated up to $T=12$ eV$^{-1}$.
We show the results in Fig.~\ref{SiH4_RT} (a).
The bound excited states is reasonably described in the calculation.
On the other hand, the calculation without an absorbing potential
(thin line) apparently provides a wrong response
in the continuum region.
We must adopt an imaginary potential to remove several spurious resonances.
However, it is very difficult to treat the ionization energy properly.
In order to mimic the continuum, the absorber should be effective only
for a outgoing electron with positive energy.
The problem is that the static potential has $-e^2/r$ behavior
at large $r$.
Electrons with negative energies ($-e^2/R<E<0$) may be
absorbed by the imaginary potential.
Therefore, the effective ionization potential becomes 
$-\epsilon_{\rm HOMO}-e^2/R$ for this calculation.
Taking $R=7$ \AA, this shift in the ionization potential amounts to
about 2 eV.
The calculated peak in the continuum is located at $\omega=14$ eV.
According to the condition (\ref{absorb_cond_1}) with $E=14-12.4+2=3.6$ eV,
we adopt the $\Delta r=10$ \AA\ and $W_0=4$ eV.
The number of grid points is 321,781 for this case.
The thick line in Fig. \ref{SiH4_RT} (a) shows the result.
The spurious peaks disappear and
the result well agrees with that of the Green's function method.
The oscillator strengths near the ionization threshold
($11<\omega<13$ eV) indicate some discrepancies,
which we naturally expect from the above argument.
In Fig. \ref{SiH4_RT} (b), integrated oscillator strength is plotted
against the energy. Seven out of eight (TRK sum rule) unit of strength 
locates below 30 eV.

Finally we should mention the feasibility of computation.
In order to calculate the absorption spectra of Fig. \ref{SiH4_RT},
the real-time method takes about 10 hours using a single CPU of
a Fujitsu VPP700E.
On the other hand, the Green's function method takes about 30
minutes for each energy.
Thus, the real-time method is faster than the Green's function method
if the response is calculated for over 20 different energies.
We have carried out the calculations for 125 frequencies to obtain
the smooth line of Fig. \ref{SiH4_GF}.

\subsection{Valence shell photoabsorption of acetylene}
\label{sec: C2H2}

The acetylene molecule, C$_2$H$_2$,
has a symmetry configuration of $D_{\infty h}$.
This high symmetry has made possible a calculation of the Green's function
using a single-center expansion \cite{LS84}.
Even so, only two kinds of molecular orbitals, $3\sigma_g$ and $1\pi_u$,
which are primarily derived from
atomic $p$ states have been taken into account in Ref. \cite{LS84},
because it was difficult to describe the $s$-derived states in the
single-center formulation.
In the present paper, we consider all valence orbitals, including
the $2\sigma_g$ and $2\sigma_u$ in addition to the above $p$-derived orbitals,
to calculate the photoresponses.

The spherical box is taken as $R=6$ \AA\ with meshes of
$\Delta x=\Delta y=\Delta z=0.3$ \AA.
All atoms are located on the $z$-axis at $\pm 0.601$ \AA\ for carbon
and $\pm 1.663$ \AA\ for hydrogen.
There are ten valence electrons and
the calculated energies of occupied orbitals are listed in Table \ref{IP}.
Using the Green's function method,
we obtain the photoabsorption oscillator strengths
as a function of photon energy, shown in Fig. \ref{C2H2_GF}.
The calculation indicates a sharp bound resonance at $\omega=9.6$ eV
and a broad structure around 15 eV which seems to be a superposition of
three resonances.
The resonance at 9.6 eV strongly responds to a dipole field parallel
to the molecular axis.
The large oscillator strengths in the IPA at $\omega=5\sim 8$ eV and
at $\omega=12.5\sim 13.5$ eV
are shifted to higher energies by the dielectric effects.
The agreements with the experimental data are significantly improved
by the inclusion of the dynamical screening.
The static dipole polarizability is also affected significantly:
In the IPA calculation,
the polarizabilities parallel ($\alpha_\parallel$) and perpendicular
($\alpha_\perp$) to the molecular axis are $\alpha_\parallel=10.7$ \AA$^3$
and $\alpha_\perp=3.87$ \AA$^3$.
The dynamic screening reduces these values to 4.79 \AA$^3$ and 2.77 \AA$^3$,
respectively, which well agree with the experimental values,
$\alpha_\parallel=4.73$ and $\alpha_\perp=2.87$ \AA$^3$ \cite{BB66}.

We find some disagreements between the calculation and the experiment in
Fig. \ref{C2H2_GF}.
We observe two distinct peaks in the experiments for the broad resonance
around 15 eV.
However, the calculation indicates three peaks.
The lowest (13.2 eV) and the highest ones (15.9 eV) are related with
responses to a dipole field parallel to the molecular axis,
while the middle one (14.3 eV) is a response to the perpendicular field.
We plot the partial oscillator strengths in an inset of Fig. \ref{C2H2_GF}.
The lowest peak at 13.2 eV consists of the transition of $3\sigma_g$
valence orbitals. The middle peak at 14.3 eV turns out to consist of 
contributions of $2\sigma_u$ and $1\pi_u$ orbitals.
We need an energy shift of these strengths by about 1 eV to reproduce
the experiments.
An accurate configuration interaction calculation with the Schwinger
variational method is available for this molecule\cite{WL99}.
This calculation succeeds to reproduce the double peak structure.
The assignments are consistent to ours: $3\sigma_g$ for lower and
$2\sigma_u$ for higher transitions.
\begin{minipage}{0.45\textwidth}
\begin{figure}[htb]
\centerline{\includegraphics[width=0.9\textwidth]{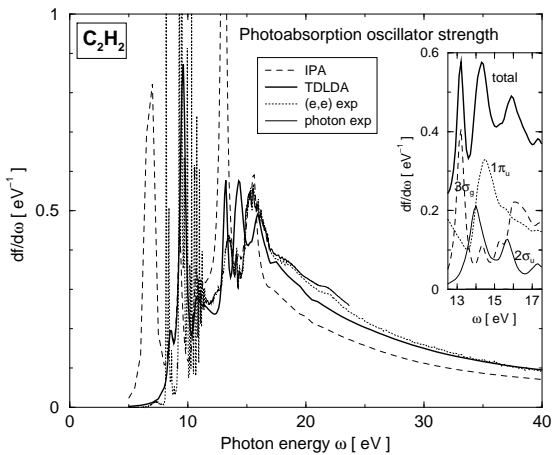}}
\caption{
Calculated and experimental photoabsorption oscillator strengths
of acetylene.
See the caption of Fig. \ref{SiH4_GF}.
The experimental data are taken from Ref. \protect\cite{Uka91,CBB95}.
Inset: An energy region of $12.5<\omega<17.5$ eV is magnified and
the total oscillator strength (thick line)
is decomposed into partial oscillator
strengths associated with different occupied valence orbital,
$2\sigma_u$ (solid line), $3\sigma_g$ (dashed)
and $1\pi_u$ (dotted).
The contributions of $2\sigma_g$ electrons are negligible.
}
\label{C2H2_GF}
\end{figure}
\end{minipage}

\begin{minipage}{0.45\textwidth}
\begin{figure}[htb]
\centerline{\includegraphics[width=0.9\textwidth]{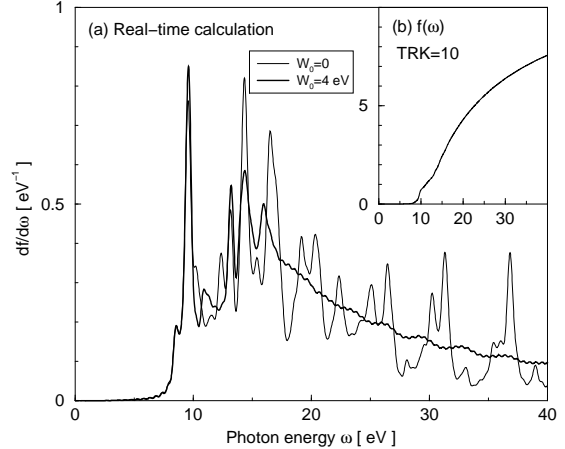}}
\caption{
(a) Photoabsorption oscillator strengths of acetylene calculated with
the real-time method.
The thin line corresponds to the result obtained with the box of $R=10$ \AA\ 
without a complex absorbing potential, while
the thick line to the one with a complex potential
of $W_0=4$ eV and $\Delta r=10$ \AA\ in addition to a spherical box
of $R=6$ \AA.
In carrying out the Fourier transform, the smoothing parameter $\Gamma=0.5$
eV is used.
(b) The integrated oscillator strength $f(\omega)$.
The TRK sum rule indicates $f(\infty)=10$ for the valence shell photoabsorption.
}
\label{C2H2_RT}
\end{figure}
\end{minipage}
\bigskip

The experiments also indicate discrete Rydberg series around 10 eV.
We do not see this structure in the calculation because we have used 
a smoothing parameter $\Gamma=0.5$ eV which is much larger than the
experimental energy resolution.
In principle, we can calculate these Rydberg states since the potential
has a $-e^2/r$ behavior at large distance.
In order to obtain these fine structures,
we have to calculate the response with $\Gamma\approx 0$
and the small frequency
mesh $\Delta\omega$ (order of meV) must be adopted.
This is beyond a scope of this paper.

The real-time calculation is carried out with the same imaginary potential
as we have used for silane ($W_0=4$ eV and $\Delta r =10$ \AA).
We have 635,371 grid points in this space.
The time evolution is calculated up to $T=12$ eV$^{-1}$ with a time step
of $\Delta t=0.001$ eV$^{-1}$.
The results are shown in Fig. \ref{C2H2_RT} (a).
Again the absorbing potential is an essential ingredient to obtain
sensible results in the continuum energy region.
The calculated photoabsorption spectra well agree with those of the
Green's function calculation except for minor oscillatory behaviors at
high energy.
This spurious oscillations start to appear at an energy
$\omega\approx 22$ eV.
In fact,
the condition of a good absorber, Eq. (\ref{absorb_cond_1}), is not
satisfied in this energy region.
Namely the potential ($W_0=4$ eV) is too week to absorb such high energy
particles.
In Fig. \ref{C2H2_RT} (b), we show the integrated oscillator strengths
up to 40 eV.
One-fourth of the TRK sum rule value of valence electrons is
in an energy region over 40 eV.
It is worth noting that
the CPU time of the real-time calculation is less than one-fifth of
that of the Green's function calculation.

\subsection{Valence shell photoabsorption of ethylene}
\label{sec: C2H4}

The ethylene C$_2$H$_4$ is the simplest organic $\pi$-system, which
possesses the $D_{2h}$ symmetry.
The two carbon atoms are on the $x$-axis at $x=\pm 0.6695$ \AA\ 
and four hydrogen atoms are in the $x-y$ plane at
$(x,y)=(1.2342,0.9288)$, $(1.2342,-0.9288)$, $(-1.2342, 0.9288)$,
$(-1.2342,-0.9288)$ in units of \AA.
There are twelve valence electrons, and
calculated eigenenergies of
occupied valence orbitals are listed in Table \ref{IP}.

The calculations are carried out using the same box ($R=6$ \AA)
and mesh size (0.3 \AA) as we have used for the acetylene molecule.
The photoabsorption oscillator strengths calculated with the
Green's function method are shown in Fig. \ref{C2H4_GF}.
The agreement with an experiment \cite{COB95} is excellent.
Almost all the main features of photoabsorption spectra
are reproduced in the calculation.
The observed bound excited states show
different photoresponses according to
the direction of the dipole field.
The lowest peak at $\omega=7.6$ eV is mainly a response to a
dipole field parallel to the molecular (C$-$C) axis.
This is associated with the excitations of the HOMO $1b_{3u}$ electrons.
On the other hand, states at 9.8 eV respond almost equally
to a dipole field of $x$, $y$, and $z$ direction,
to which both $1b_{3g}$ and $1b_{3u}$ occupied orbitals contribute.
A small peak at $\omega=11.4$ eV is calculated as a resonance of
$3a_g$ orbital, 
which may correspond to a small shoulder in the experiment.
Beyond 11.7 eV, the HOMO electrons are in the continuum.
The first prominent peak at 12.4 eV is a resonance with respect to
a dipole field of $y$ direction which is in the molecular plane
and perpendicular to the C$-$C bond.
The excitations of the $1b_{2u}$ and $1b_{3g}$ electrons are  main
components of this resonance.
In a region of $13.2<\omega<20$ eV,
the $1b_{3g}$ electrons can be excited into the continuum and
produce the smooth background of oscillator strengths
($0.1\sim 0.2$ eV$^{-1}$).
A peak structure at 14.6 eV is made of
the excitation of $1b_{2u}$ electrons.
The peak at 16.4 eV is constructed by excitations from
the $2b_{1u}$ occupied orbitals, while the experiment indicates the 
resonance at 17.1 eV.

The above analysis is consistent with that in the literature\cite{COB95},
except for the 12.4 eV peak (11.9 eV in the experiment). Our analysis
suggests transition of $1b_{3g}$ and $1b_{2u}$ electrons while 
Ref.\cite{COB95} indicates $3a_g$. 

Comparing the TDLDA calculation with that of the IPA (dashed line),
we see again that the dynamic screening effect is very important
to reproduce the experimental data.
This feature of the IPA result is consistent with the other calculations
without the dynamic screening \cite{Gri83}.
The calculated dipole polarizability is
$\alpha=4.22$ \AA$^3$ (5.47, 3.97, 3.23 \AA$^3$
for $x$, $y$, $z$ direction, respectively),
which is in a good agreement with
the experimental value, 4.22 \AA$^3$ \cite{BB66}.
In the IPA calculation, we obtain
$\alpha=7.04$ \AA$^3$ (10.4, 5.96, 4.76).

Results of the real-time calculation are shown in Fig. \ref{C2H4_RT} (a).
The box and imaginary potential we use are the same as those for acetylene.
We can obtain a sensible result if we adopt the absorbing potential.
However, again, a spurious oscillatory behavior is seen at high energy region
because the absorbing potential is not strong enough to erase those
high-energy components.

\begin{minipage}{0.45\textwidth}
\begin{figure}[htb]
\centerline{\includegraphics[width=0.9\textwidth]{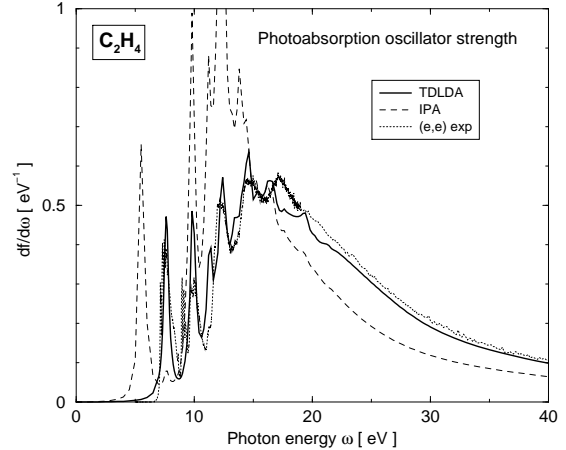}}
\caption{
Calculated and experimental photoabsorption oscillator strengths
of ethylene.
The thick solid line is the calculation compared with
high resolution dipole (e,e) experiment (dotted line) \protect\cite{COB95}.
The dashed line is the IPA calculation without dynamical screening.
The smoothing parameter $\Gamma=0.5$ eV is used in the calculation.
}
\label{C2H4_GF}
\end{figure}
\end{minipage}

\begin{minipage}{0.45\textwidth}
\begin{figure}[htb]
\centerline{\includegraphics[width=0.9\textwidth]{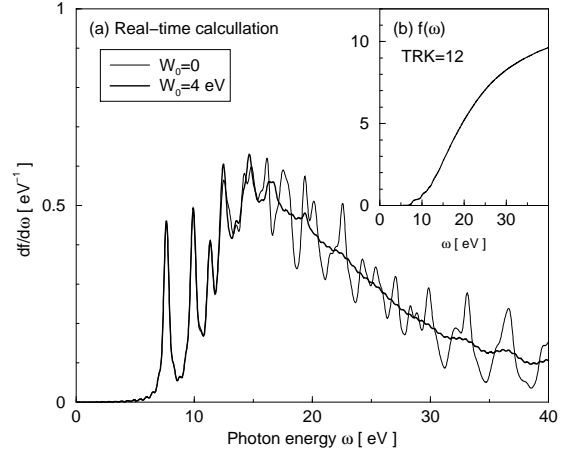}}
\caption{
Photoabsorption oscillator strengths of ethylene calculated with
the real-time method.
See the caption of Fig. \ref{C2H2_RT} (a).
(b) The integrated oscillator strength $f(\omega)$.
The TRK sum rule indicates $f(\infty)=12$ for the valence shell photoabsorption.
}
\label{C2H4_RT}
\end{figure}
\end{minipage}

\section{Conclusions}
\label{sec: conclusions}

We have developed methods based on the TDLDA of investigating
the photoresponses in the continuum for systems with no spatial symmetry:
(1) The real-time method with an absorbing potential, and
(2) the Green's function method.
These methods allow us to treat the photoionization and the dynamical
screening effects self-consistently.
For the real-time method, we have tested imaginary potentials of different
kinds and found that those satisfying a condition,
Eq. (\ref{absorb_cond_1}), are good to mimic the continuum effect.
However, it is very difficult to treat the photoabsorption spectra in
vicinity of the ionization threshold.
There are two reasons: One is because the condition,
Eq. (\ref{absorb_cond_1}), requires a large model space to treat such a
low-energy emission properly.
The second is that the ionization threshold is not correct when the
potential has a $1/r$ behavior at large distance.
In addition to this, since the condition is energy dependent,
it is very difficult to construct a good absorber for low-energy and
high-energy outgoing electrons simultaneously.
The advantage of the real-time method is a computational feasibility.
Utilizing the Fourier transform, we can obtain the spectra over the whole
energy region at once.

The Green's function method possesses a capability
of an exact treatment of the continuum.
Using a Green's function of Coulomb asymptotic waves,
it is also possible to investigate the photoresponse near the threshold.
The main difficulty of the Green's function treatment is a heavy
computational task.
In order to reduce the CPU time, we use a complex energy for the
Green's function, $G(\omega+i\Gamma/2)$.
We have found that the inclusion of the imaginary part $\Gamma$ facilitates
a convergence of iterative procedures in the calculation.
The $\Gamma$ also plays a role of lowering an energy resolution of
the calculations.
Thus, we can choose a value of $\Gamma$ depending on the energy resolution
required in each problem.
We would like to emphasize again that the method is capable of calculating
response functions of many-electron systems,
below, near and above the ionization threshold in a unified manner.

The numerical calculations have been performed for a spherical Na$_7^-$ cluster
(test study) and non-spherical molecules, SiH$_4$, C$_2$H$_2$, and C$_2$H$_4$.
Studies of photoresponse for deformed molecules
including both the dynamic screening and the continuum effect are very few.
An exception is a study of nitrogen and acetylene molecules
(with a high symmetry $D_{\infty h}$) by Levine and Soven \cite{LS84}
using the single-center expansion technique.
However, only $1\pi_u$ and $3\sigma_g$ occupied orbitals are included in
their calculation because of limitation of the single-center expansion.
Since our calculation has been carried out on three-dimensional coordinate
meshes, we do not need any spatial symmetry including all valence orbitals
in the calculation.
We present the photoabsorption oscillator strengths compared with
dipole $(e,e)$ and synchrotron radiation experiments.
The agreement is generally excellent.
The inclusion of the dynamic screening turns out to be essential to
reproduce the experiments.
The IPA calculation overestimates the strengths at low energy and
underestimates at high energy.

We are strongly encouraged by the success of our methods applied to
simple molecules in the present paper.
The applications to more complex systems may
become possible in near future.

\acknowledgments

We would like to thank Y.~Hatano and K.~Kameta for useful discussion
and providing us with the photoabsorption experimental data.
Calculations were performed on a Fujitsu VPP700E Super Computer at
RIKEN, a NEC SX-4 Super Computer at Osaka University
and a HITACHI SR8000 at Institute of Solid State Physics,
University of Tokyo.


\end{multicols}

\end{document}